# Apply VGGNet-based deep learning model of vibration data for prediction model of gravity acceleration equipment

SeonWoo Lee[a,1], HyeonTak Yu[b,2], HoJun Yang[a], JaeHeung Yang[c], GangMin Lim[c], KyuSung Kim[d], ByeongKeun Choi[b,+] and JangWoo Kwon[a*]

[a]*Deparment Electric Computer Engineering, Inha University, 100, Inha-ro, Michuhol-gu, Incheon, Republic of Korea*
[b]*Department Mechanical Engineering, Gyeong-Sang National University, 38, Cheondaegukchi-gil, Tongyeong-si, Gyeong sangnam-do, Republic of Korea, 530-64*
[c]*R&D Center, ATG, Seongnam-daero, Bundang-gu, Seongnam-si, Gyeonggi-do, Korea*
[d]*Department of Otolaryngology-Head and Neck Surgery, Inha University College of Medicine, Incheon, 3-Ga Shinheungdong, Jung-Gu, Incheon 400-711, Korea*

**Abstract.** Hypergravity accelerators are a type of large machinery used for gravity training or medical research. A failure of such large equipment can be a serious problem in terms of safety or costs. This paper proposes a prediction model that can proactively prevent failures that may occur in a hypergravity accelerator. The method proposed in this paper was to convert vibration signals to spectograms and perform classification training using a deep learning model. An experiment was conducted to evaluate the performance of the method proposed in this paper. A 4-channel accelerometer was attached to the bearing housing, which is a rotor, and time-amplitude data were obtained from the measured values by sampling. The data were converted to a two-dimensional spectrogram, and classification training was performed using a deep learning model for four conditions of the equipment: Unbalance, Misalignment, Shaft Rubbing, and Normal. The experimental results showed that the proposed method had a 99.5% F1-Score, which was up to 23% higher than the 76.25% for existing feature-based learning models.

## 1. Introduction

All objects on Earth are affected by the Earth′s gravity. Conducting research on microgravity on the ground, instead of outer space, has many practical difficulties. On the other hand, research on hypergravity is relatively easy to carry out using the centrifugal force from a spinning simulation.

---

[*]Corresponding author. E-mail: jwkwon@inha.ac.kr
[+]Corresponding author. E-mail: bgchoi@gnu.ac.kr

Hypergravity research requires a gravity simulator that can control gravity by a constant rotation angular speed. Therefore, to conduct hypergravity research, a gravity simulator was developed to enable the formation and maintenance of a hypergravity environment of up to 15 times the Earth′s gravity (15 G), as shown in Fig 1.

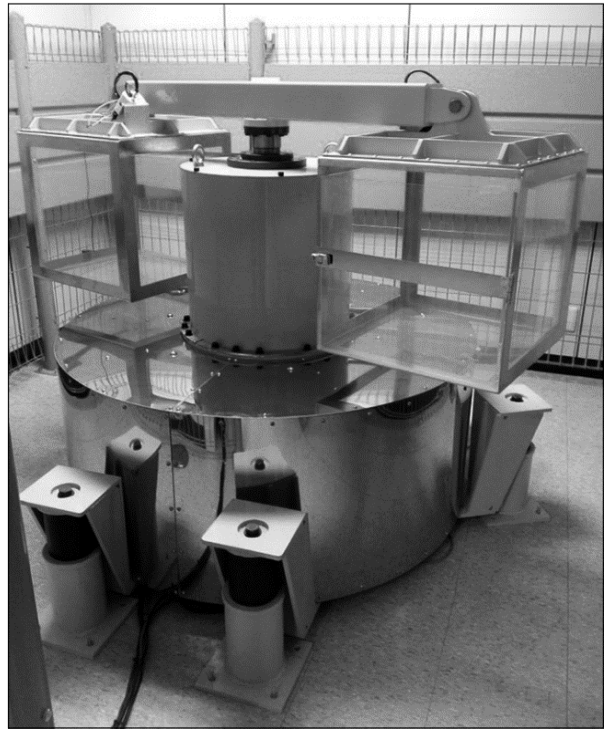

Fig. 1 Gravity Simulator for the research on hypergravity [23]

Gravitational accelerators are generally used for the hypergravity training of astronauts and can be used for animal testing in basic research for medical purposes. In addition, they can be used to conduct experimental ground tests on the effects of sudden changes in gravity, such as hypergravity and hypogravity, and the changes in pressure that the human body undergoes in a space environment to investigate the biological responses to these harmful stimuli to the human body. These changes in gravity can result in fluid shifts and redistribution in the human body, fluid loss, red blood cell loss, muscle damage, bone damage, hypercalcemia, immune system changes, or spatial disorientation and vertigo [1].

Many studies have examined the changes in the human and animal body due to changes in gravity. The necessity of monitoring the safety and reliability of large gravity acceleration equipment has become an important issue. One of the major issues regarding gravity acceleration equipment is the occurrence of abnormal vibrations when machinery failures occur due to high-speed rotation. The amplification of small vibrations generated in the rotating part of the gravity acceleration equipment may result in damage to the shafts rotating at high speeds, which may lead to serious accidents. Therefore, this study aimed to detect abnormal vibrations of gravity acceleration equipment.

Many studies on vibration-related failures and predictive failure diagnosis have been conducted.[2, 3, 4, 5, 6] The major methods used in these studies [6] were characterized largely by the application of algorithms, such as SVM (Support Vector Machine)[7] after calculation and enumeration of the features, such as the mean, standard deviation, fast Fourier transform (fft), and kurtosis and selection of features that appropriately express the patterns or properties for classification tasks based on the genetic algorithm or principal component analysis. In recent years, ANNs (Artificial Neural Networks) have become one of the most widely used methods because of the advances in hardware technology and a large number of data that express the relevant features well. [8]. In the case of the CNN (Convolution Neural Network), which is an ANN method, training is carried out using the following procedure: multiple inputs are received; the computation is performed using a model form that the user wants; an output is produced. The method of applying a 1-D CNN model using time-amplitude data with a constant period has been presented as a failure diagnosis method [9, 10]. Another CNN model is 2-D CNN, in which the computation produces images of 3-D shapes with a width and length like the input data as the output [11, 12, 13]. Many attempts have been made to apply 2-D CNN to speech recognition and failure diagnosis because of its high performance [14, 15, 16, 17].

This paper proposes a preventive maintenance model that enables the monitoring of vibrations that can occur in machinery to prevent proactively the mechanical failures described above. Traditional machine learning methods that use feature-based methods [6, 7, 8] based on human-designed lists of feature engineering have limitations that can improve performance significantly. Therefore, CNN was used to maximize the performance.

The method proposed in this study was divided into two major methods. The first method was to convert vibration data into two-dimensional data by converting time-amplitude data with a constant period, such as the existing signals, into spectrograms. These spectrograms display the time, frequency (Hz), and amplitude, which are used mainly for speech recognition [14, 16, 17]. The second method was to apply the preprocessed data to a deep neural network model and compare the results with those obtained by the existing machine learning models.

The remainder of this paper is organized as follows. Section 2 briefly introduces the proposed method and the equipment used in the experiment. We try to show the excellence of the method provided through comparative analysis with conventional research in Section 3. In Section 4, experimental results are presented, and Section 5 reports the conclusions of this study and the direction of future research.

## 2. Proposed Method and Environment

### 2.1. Design and fabrication of experimental rotating equipment

The simulation equipment was manufactured as described below. Pulse 3560C and four accelerometers (B&K 4371) were used to acquire the rotation and vibration data, and the data acquisition time for each condition was 30 seconds. Table 1 lists the specifications of the data acquisition system.

Table 1 Properties of the data acquisition system

| Type | Properties |
|---|---|
| Pulse 3560C (B&K) | 4/2-ch Input/output Module<br>Operating Freq. range: 0~25.6kHz<br>Direct/CCLD/MIC. preamp 1 Tacho<br>Conditioning |

| Accelerometer (B&K 4371) | Operating Freq, range: 1~25.6kHz<br>Operating Temp. -50C~121C<br>Sensitivity : 9.84 pC/g |
|---|---|

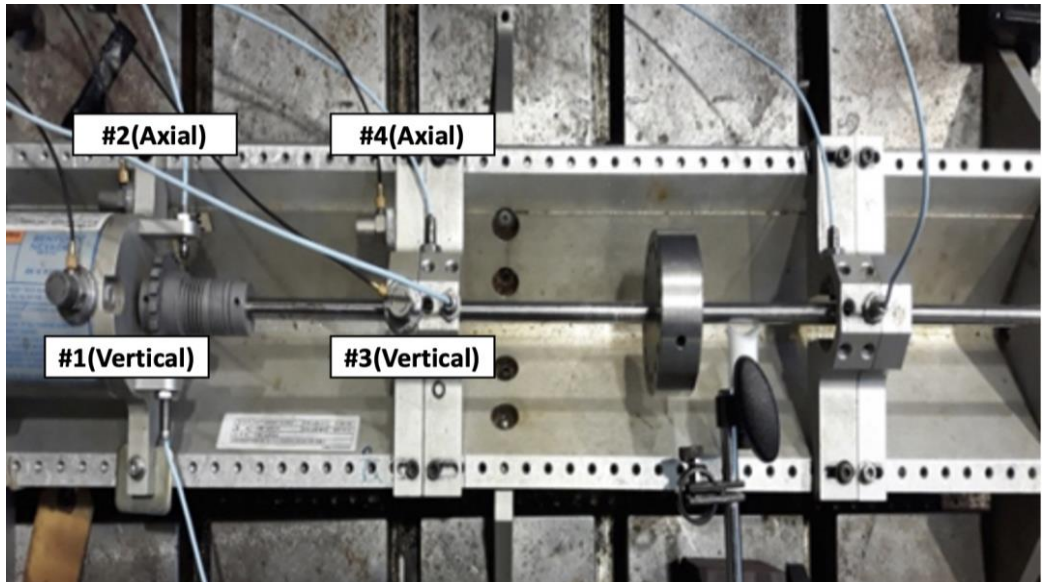

Fig. 2 Measurement of experiment system

Fig.2 presents the RK4 (Rotor-kit) of the lab-scale rotating simulation equipment, which is the experimental model, and the locations of the sensors used in the experiment. The experiment system was composed of a motor to operate the rotating equipment, a flexible coupling connecting the rotor and motor, and two copper sleeve bearings supporting the rotor. An 800g disk was installed between the bearings to simulate the unbalance fault. Sensors were installed on the drive-end side of the motor and rotor. The measurements were taken at locations in the vertical and axial directions of the motor and rotor. The experimental equipment was operated at 2000 RPM, avoiding 2400 RPM, which is the first critical speed.

In this experiment, fault simulations were carried out by simulating four representative conditions of the rotating equipment: Normal, Unbalance, Misalignment, and Shaft rubbing conditions. Fig. 2 presents the methods of application of the Normal condition and each type of fault. A normal condition was obtained after performing shaft balancing using the RK4, and the residual unbalance was measured to be 0.02g/117.4° after balancing. Unbalance was induced by attaching a 3.2 g object in a direction towards the location of residual unbalance (117.4°). Misalignment was achieved by installing a 4mm shim plate at the foot of the drive-end side of the motor, and shaft rubbing was applied in the horizontal direction using a magnetic base. In addition, a contact device made from Teflon was used to minimize the damage to the axis that may occur due to rubbing.

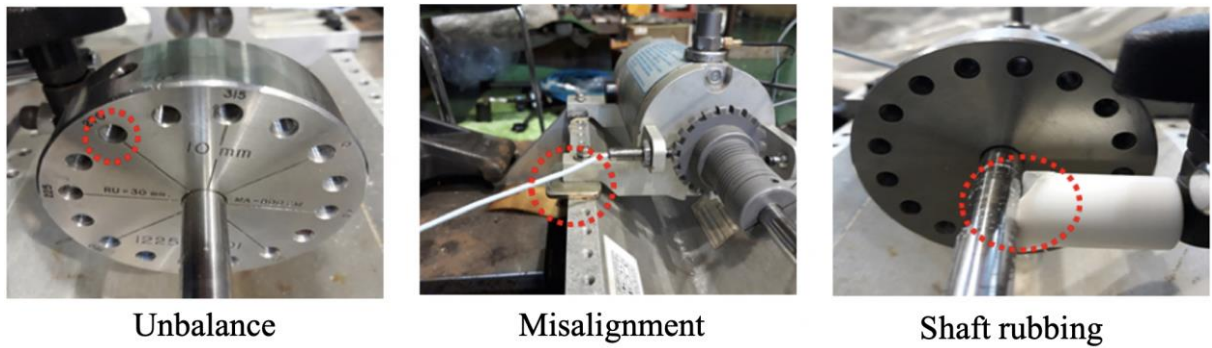

Fig. 3 Measurement of the experiment system

Unbalance is the most fundamental fault that causes vibrations in rotating equipment. Unbalance occurs when the mass distribution of the rotor is asymmetric with respect to the axis centerline, and all the causes of unbalance exist to some degree in the rotors. Excessive unbalance increases the vibrations and noise of the rotating equipment. As a result, fatigue destruction may occur due to a deterioration of the bearings and consumable parts.

Misalignment is one of the most common faults of rotating equipment along with unbalancing [18] and refers to a condition where the centers of the two axes do not coincide, or a condition where the centers coincide but are not parallel. A large degree of misalignment can cause overheating of the coupling, an increase in the shaft cracks and fatigue, and damage to the bearings and consumable parts.

A rubbing fault is a secondary transient phenomenon caused by excessive unbalance and misalignment in rotating machinery [19]. Rubbing may be caused by the occurrence of friction between the stator and rotor caused by excessive vibrations, or a narrow gap due to thermal expansion during equipment operation. Continuous rubbing during the operation of rotating machinery may cause the separation of parts or axis bending, and severe rubbing can lead to the destruction of the rotating equipment.

The sampling rate of the obtained signals was 65536 Hz. The signals measured for 30 seconds were divided into 0.48-second units considering the measurement environment of the actual equipment, and each of the 0.48-second units was assumed to be one data set. Machine learning was performed by dividing one data set into 14 samples. Sampling was performed because a vibration is a periodic signal in the time domain [20], and most fault signals have periodicity. Therefore, sampling is used to examine the consistency and continuity of each condition using the features calculated from the signals.

The signal segmentation for sampling was based on the rotational frequency of the rotor. Generally, in rotating equipment, the rotational frequency is the most dominant component, and the majority of fault components appear in the harmonic form of the rotational frequency. Therefore, the length of the sample of experimental data was set to 0.06 seconds. This was two times 0.03 seconds, which is the period of vibrations at 2000 RPM, and the number of samples was increased by overlapping half the signal.

The total number of training and test data was 1056, and the dataset was divided into training and testing datasets by allocating 80% to the training dataset and 20% to the testing dataset. At this time, the training dataset consisted of 229 Normal data, 199 Rubbing data, 205 Unbalance data, and 211 Misalignment data, and the testing dataset included 43 Normal data, 61 Rubbing data, 55 Unbalance data, and 53 Misalignment data.

*2.2. Method of data visualization*

Figure 4 shows the data conversion method proposed in this paper. After receiving data from the experimental equipment at 0.06-second intervals, as described in Section 1.1, the data were converted to STFT (Short Time Fourier Transform) signals or MFCC (Mel Frequency Cepstral Coefficient) signals and converted again to spectrograms. Through this process of data visualization, the vibration data were converted to spectrograms.

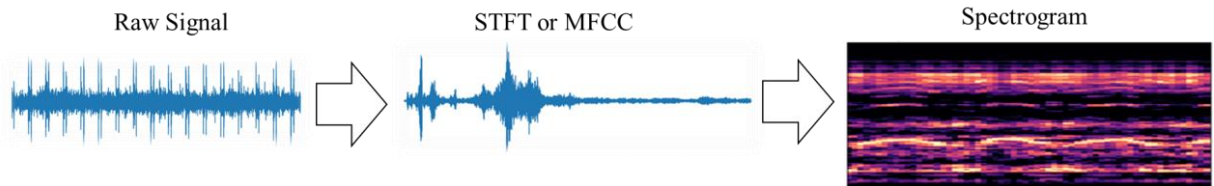

Fig. 4 Example of the conversion of vibration signals into images

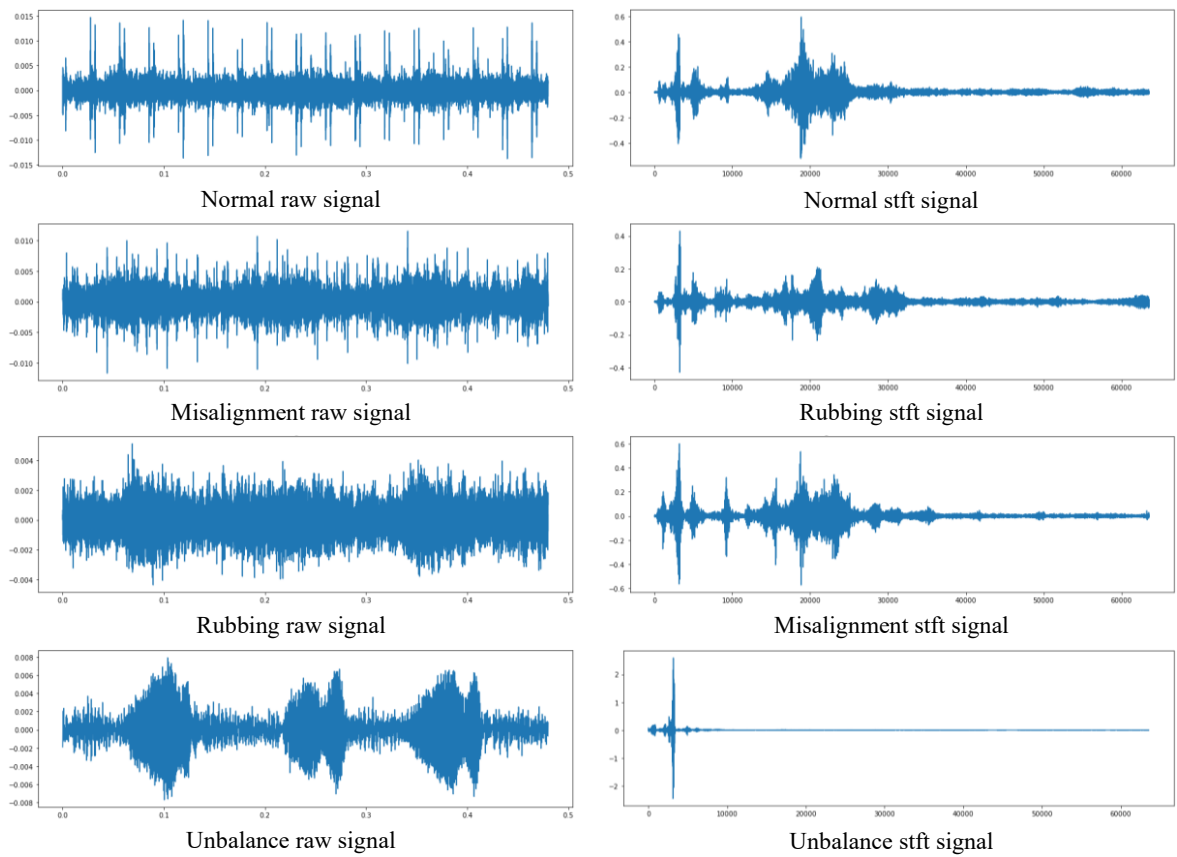

Fig. 5 Original signals of each target class: Normal, Misalignment, Rubbing, and Unbalance (Left); Examples of SFTF conversion of Normal, Misalignment, Rubbing, and Unbalance signals (Right)

Fig. 5 gives an example of the representation of signals for the Normal condition and each type of Fault described in Section 1.1 as two-dimensional amplitude-time graphs and an example of converting the graphs using the STFT. In general, a fault signal tends to have a larger amplitude than a normal signal.

*2.3.1 STFT (Short-Time Fourier Transform)*

With respect to the method for converting time-amplitude data to 2D images, spectrograms were used after performing discrete STFT. The discrete STFT is a method of partitioning continuous signals over a long period into shorter segments at short time intervals and applying a Fourier transform to each signal segment. This technique allows researchers to observe how the vibrations of signals change with time. These changes in vibrations can be expressed as Eq. (1) [24][25]

$$X(k,n) := \sum_{m=0}^{L-1} \omega[m]x[m+nH]exp(-2\pi k/N)m \quad (1)$$

$\omega[m]$ was assumed to be a non-zero window function in the interval $m = 0,1,\cdots,L-1$, and $L$ is the window length, and a smaller signal than the signal $x[m]$. In this experiment, the Han window was applied as the window function [25]. $\omega[m]x[m+nH]$ is a non-zero signal in $m = 0,1,\cdots,L-1$. The signal $x[m]$ is a form that undergoes $N$ point DFT (Discrete Fourier Transform) according to the hop size of $H(=512)$. The hop size $H$ is specified in samples and determines the step size moving through the window in the overall signal. [25] Therefore, FFT was calculated according to the size of $m$. Because a signal generated through this process constitutes a different spectrum with time, it cannot be represented as a spectrum. Therefore, it was represented by taking $|X(k,n)|$ and applying a color map(Spectrogram), as shown in Fig.6.

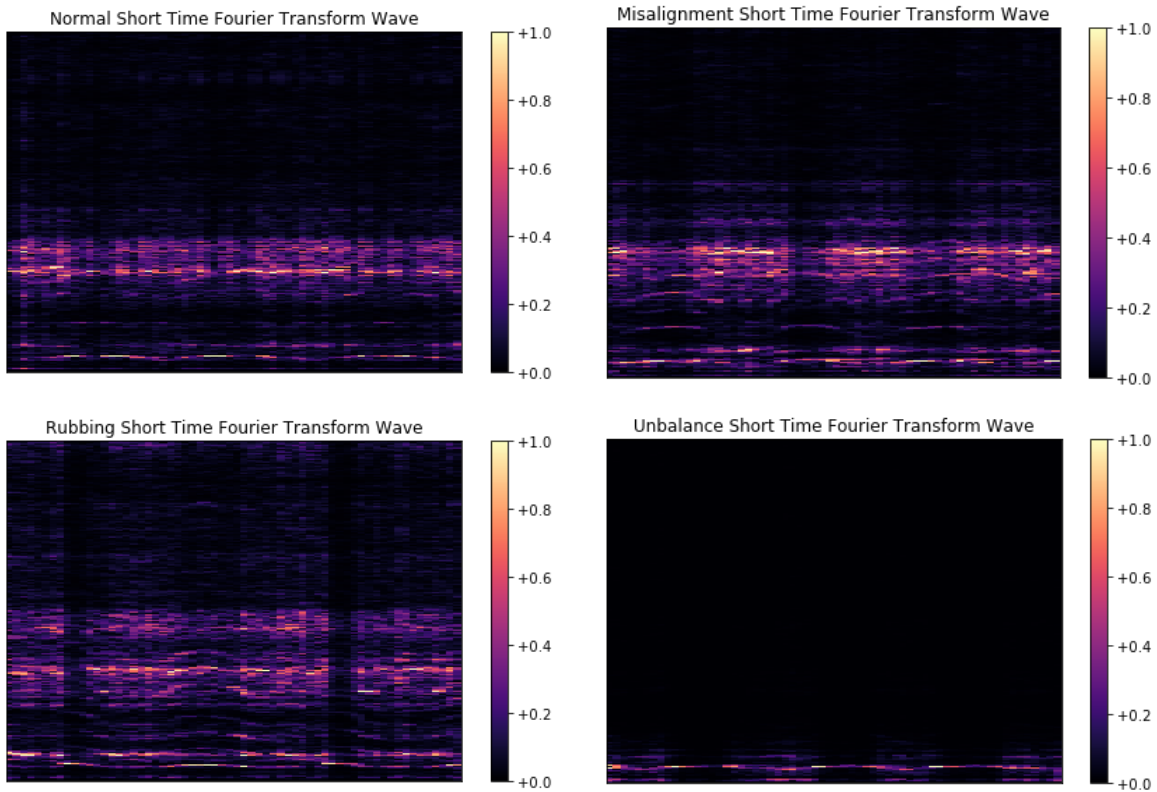

Fig. 6 Examples of the changes in STFT spectrograms

*2.3.2 MFCCs(Mel Frequency Cepstral Coefficients)*

MFCC is a conversion algorithm used mainly in speech recognition. This is one of the methods for extracting the features from sound signals and the procedure for feature extraction consists of the following six steps [16, 17]:

- Frame the signal into short frames.
- For each frame, calculate the periodogram estimate of the power spectrum.
- Apply the mel filterbank to the power spectra, and sum the energy in each filter.
- Take the logarithm of all filterbank energies.
- Take the DCT of the log filterbank energies.
- Keep DCT coefficients 2-13, and discard the rest.

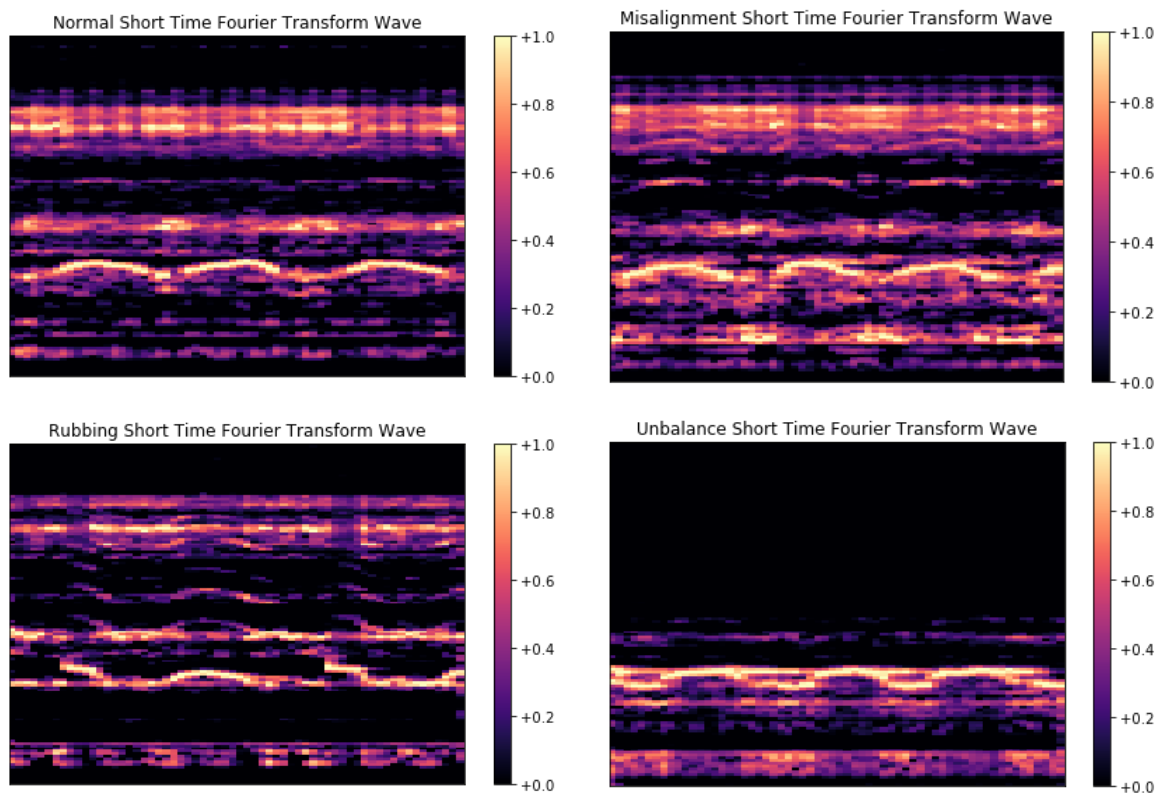

Fig. 7 Example of signal changes in the spectrograms of the MFCC signals

*2.3. Deep learning network*

The deep learning neural network architecture proposed in this study was based on VGG19 [11]. VGG19 is a model that is used widely as a basic deep learning method because it is relatively easy to implement and modify because it uses only $3 \times 3$ convolutional layers. In this study, the number of parameters was reduced using ′average pooling′ to eliminate the last′ fully connected layer′, which is one of the parts of VGGNet that requires a large number of computations, and to match with the output layer. The deep learning architecture was constructed, as shown in Fig. 4.

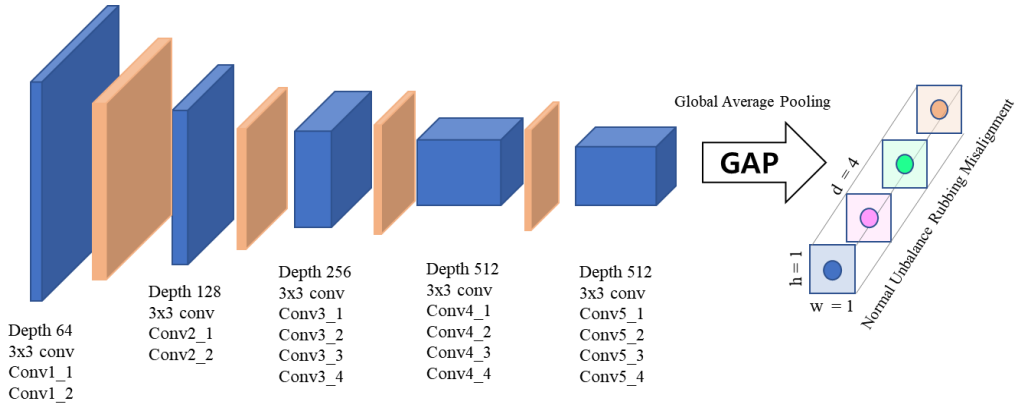

Fig. 8 Proposed network architecture. The amount of computation was reduced by replacing the last Fully Connected Layer with Global Average Pooling.

The image size of the spectrogram in Fig. 8 used as a training data was changed by converting a rectangular shape (432, 288) to a square shape (298, 298) before using it in the experiment. For convergence of the learning errors, an attempt was made to find the Global Minimum Error using the Learning Rate Decay Strategy, which reduces the learning rate by 0.04% at 60, 120, and 160 epochs. The other hyperparameters were set, as shown in Table 2.

Table 2 Hyperparameters used in model training

| HyperParameter | Value |
| --- | --- |
| Learning Rate | 0.1 |
| Batch Size | 4 |
| Warm-up Train phase | 10 |
| Weight Decay | 0.04 |
| Optimizer | Stochastic Gradient Descent |
| Epoch | 200 |

An initial learning rate of 0.1 was set for faster training speeds. A batch size of four was used to set the maximum batch size in the environment to speed up learning. The early 10 epochs were used to warm-up the training phase and adjust the learning rate according to the complexity of the training data. The first 200 epochs were used for a more robust model.

## 2.4. Deep learning environment

In this study, the deep learning environment for training and testing a deep learning model was built with a PC with the following configuration: 32GB RAM, i5-8500 3.0GHz CPU, and RTX 2080 Ti GPU. The experimental software environment was developed in a Python 3.7.6 environment, and the main packages used to set up the environment were Pytorch 1.5 [20], librosa 0.6.3 [21], and sklearn 0.22 [22].

## 3. Performance evaluation

In the experiments of this study, the accuracy, precision, recall, and F1-Score were measured using True Positive (TP), True Negative (TN), False Positive (FP), and False Negative (FN). The accuracy, precision, recall, and F1 Score can be expressed using Eqs. (2) – (5), respectively.

$$Accuracy = \frac{TP+TN}{TP+TN+FP+FN} \quad (2)$$

$$Presicion = \frac{TP}{TP+FP} \quad (3)$$

$$Recall = \frac{TP}{TP+FN} \quad (4)$$

$$F1\ Score = 2 \times \frac{Precision \times Recall}{Precision+Recall} \quad (5)$$

At this time, to demonstrate the superiority of the methods used in this experiment, they were compared with one of the most commonly used methods, the method of applying SVM (Support Vector Machine), after feature selection based on the genetic algorithm after extracting the features from a raw signal and [6]. The [6] method is a method of applying SVM by mixing a GA(Genetic Algorithm) and PCA(Principal Component Analysis) from a list of 30 feature values through feature engineering. The proposed methods were also compared with the MLP (Multi-Layer Perceptron) method [8] from the same feature engineering [6] to determine if it shows better performance in training after data visualization. Table 3 lists the experimental results. Under Normal, Rubbing, Unbalance, and Misalignment conditions, the proposed methods showed better performance than the existing methods [6, 7. 8]. In this study, an attempt was made to improve performance through k-fold cross-validation, but the following problems were encountered. First, the accuracy was low compared to the results not applied because the number of datasets was not large. Second, the experimental results of the current dataset were not applied because they were unnecessary owing to the very high accuracy.

Table 3 Experimental Results

| Class | Model | Accuracy | Precision | Recall | F1 |
|---|---|---|---|---|---|
| Normal | SFTF-Spectrogram VGG19 (our) | **1.0** | 0.98 | **0.99** | **0.99** |
| | MFCC-Spectrogram VGG19 (our) | **1.0** | **1.0** | 0.98 | **0.99** |
| | MLP[8] | 0.93 | 0.95 | 0.95 | 0.95 |
| | GA-SVM[6] | 0.76 | 0.90 | 0.94 | 0.92 |
| Rubbing | SFTF-Spectrogram VGG19 (our) | **1.0** | **1.0** | **1.0** | **1.0** |
| | MFCC-Spectrogram VGG19 (our) | **1.0** | **1.0** | **1.0** | **1.0** |
| | MLP[8] | 0.93 | 0.90 | 0.89 | 0.90 |
| | GA-SVM[6] | 0.76 | 0.72 | 0.66 | 0.69 |
| Unbalance | SFTF-Spectrogram VGG19 (our) | **1.0** | **1.0** | 0.99 | **0.99** |
| | MFCC-Spectrogram VGG19 (our) | **1.0** | 0.98 | **1.0** | **0.99** |
| | MLP[8] | 0.93 | 0.96 | 0.95 | 0.96 |
| | GA-SVM[6] | 0.76 | 0.78 | 0.77 | 0.78 |
| Misalignment | SFTF-Spectrogram VGG19 (our) | **1.0** | **1.0** | **1.0** | **1.0** |
| | MFCC-Spectrogram VGG19 (our) | **1.0** | **1.0** | **1.0** | **1.0** |
| | MLP[6] | 0.93 | 0.89 | 0.91 | 0.90 |
| | GA-SVM[8] | 0.76 | 0.65 | 0.68 | 0.66 |

The performance of the deep learning methods was superior to that of a method based on MLP or SVM, as listed in Table 3. This can be attributed to a large amount of information that cannot be expressed as features that are lost when selecting the features of input data in the preprocessing stage. Although all 30 features were selected and learned using the MLP algorithm, a performance equal to or better than that of deep learning could not be achieved. As shown in the F1-score result in Table 3 by Eq (5), the classification balance performance of precision and recall was also excellent. In addition, the results were similar even when MFCC or STFT was selected during preprocessing. Most of the information on the label required could be acquired in the form of an input image.

Table 4 compares the training results based on the dataset that has undergone an STFT transformation with the existing deep learning model [11, 13, 26, 27]. The training hyperparameters of each model were trained under the same conditions, as listed in Table 2. All models to be compared were intended to converge more quickly by Transfer Learning, which reuses existing trained models. The proposed model was approximately 30% better than the Alexnet and VGG19 models. In the case of SqueezeNet, the result of the dataset proposed in this paper was not excellent.

Table. 4 Comparison of the training results with the existing deep learning models

| Model | Parameters | Accuracy | Precision | Recall | F1 |
|---|---|---|---|---|---|
| AlexNet[26] | 57,020,228 | 0.31 | 0.23 | 0.31 | 0.19 |
| VGG19[11] | 139,597,636 | 0.70 | 0.69 | 0.71 | 0.67 |
| SqueezeNet[27] | 737,476 | 0.25 | 0.06 | 0.25 | 0.10 |
| Our | 20,037,444 | 1.0 | 1.0 | 0.99 | 0.99 |

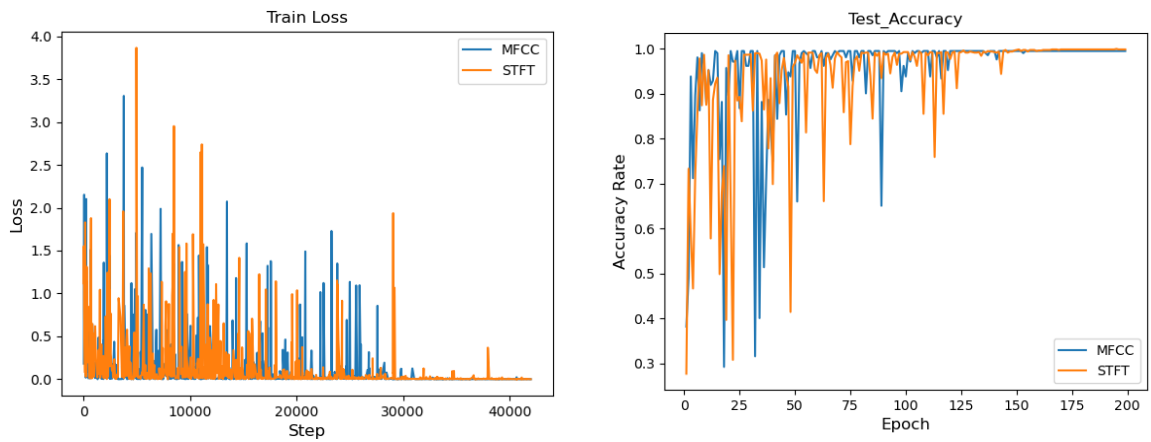

Fig. 9 Training loss (Left) and accuracy rate according to epoch number (Right) in the experiment

As shown in Fig. 9, the performance of the deep learning model is shown by the training errors and accuracy curves obtained as the experimental results. A unique aspect of this experiment was that the errors and test accuracy of the data training models to which the data preprocessed by MFCC and STFT conversions were applied, converge when the completion rate is approximately 75%. In both models, the loss value was small in the initial stages of training, which then converged as the vibration amplitude became larger. Initially, a phenomenon, in which the value of the training loss or test accuracy appeared to be unstable, occurred. This was attributed to the initial warm-up training phase.

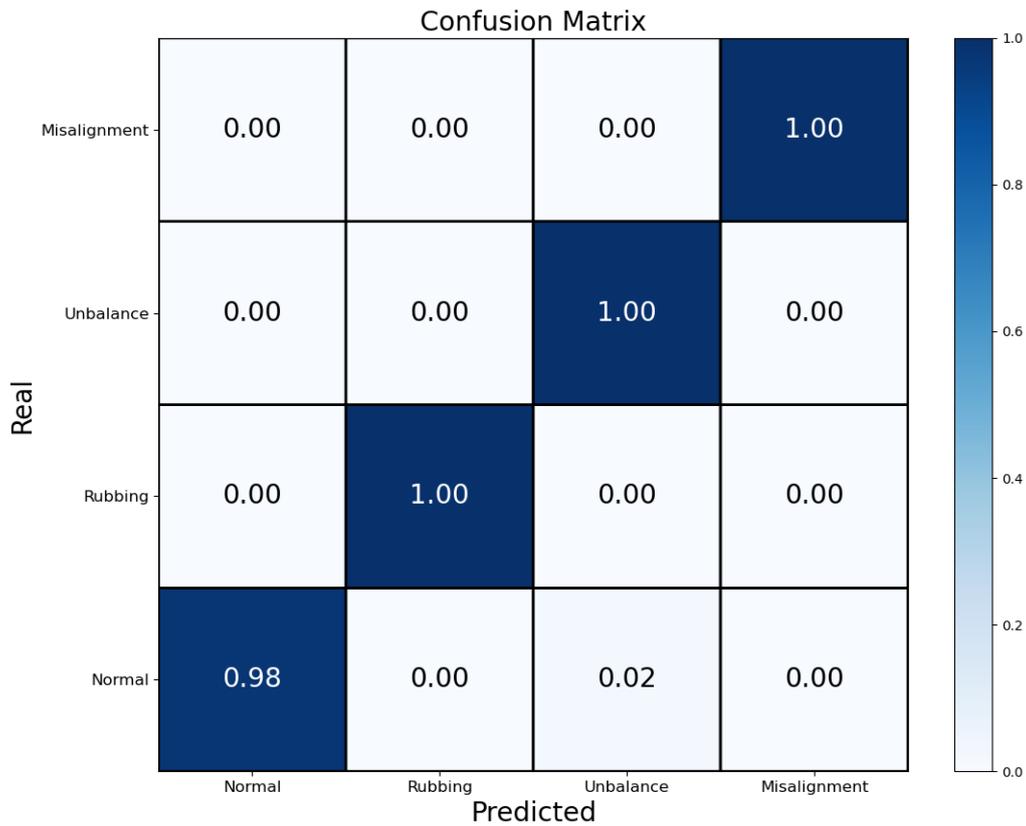
Fig. 10 Confusion Matrices obtained using MFCC and STFT. The confusion matrices of MFCC and STFT were identical.

Fig. 10 shows the Confusion Matrix of the models used in the experiments; the results of both models were the same. Regarding the training errors of the models, there was a classification error of 2% for Normal and Unbalance.

## 4. Conclusion and Future Work

The vibration signals were measured with accelerometers to prevent accidents that can occur in large equipment, such as a gravitational accelerator. In this paper, four signals that can arise when a defect occurs in the rotating part of a gravitational accelerometer were analyzed. The measured data were used to train and test a deep learning model using the spectrogram visualization based on the MFCC and STFT, and the proposed method was evaluated.

The major methods used in this experiment were to convert vibration signals to images and apply a modified VGGNetwork to a fault model. The proposed deep learning architecture enabled a diagnosis of the four conditions, such as Normal, Rubbing, Misalignment, and Unbalance. Both MFCC and STFT models showed an average accuracy of 99.5%. In addition, the proposed models were compared with feature-based machine learning models using the existing traditional methods. The experimental results showed that the proposed models have better performance in terms of accuracy, recall, precision, and F1-Score compared to existing feature-based learning models. These results suggest that the proposed method can be used successfully as a fault diagnosis and assessment model if the monitoring environment is constructed by attaching sensors in an assessment of the stability of gravity acceleration equipment in the future.

In addition, the existing vibration data can also be converted to image data, such as spectrograms, which is one of the methods used in speech recognition, and they can be applied to an image-based deep learning model. The method proposed in this study had the following limitations. The patterns of the fault data need to be prepared in advance. This should be addressed in further studies, such as research on the detection of outliers. Second, training takes considerable time and requires additional hardware, such as GPUs. Considering these limitations, a method that can reduce the computation cost so that the proposed method can be used in small edge devices will be needed before this method can be commercialized.

## 5. Acknowledgments

This study was supported by the MSIT (Ministry of Science and ICT), Korea, under the ITRC (Information Technology Research Center) support program (IITP-2017-0-01642) supervised by the IITP (Institute for Information & communications Technology Promotion)

# References

[1] Gurovsky, N. N., Gazenko, O. G., Adamovich, B. A., Ilyin, E. A., Genin, A. M., Korolkov, V. I., Shipov, A. A., Kotovskaya, A. R., Kondratyeva, V. A., Serova, L. V., & Kondratyev, Y. I. (1980). Study of physiological effects of weightlessness and artificial gravity in the flight of the biosatellite Cosmos-936. Acta Astronautica, 7(1), 113–121.

[2] Riaz, S., Elahi, H., Javaid, K., & Shahzad, T. (2017). Vibration Feature Extraction and Analysis for Fault Diagnosis of Rotating Machinery-A Literature Survey. Asia Pacific Journal of Multidisciplinary Research, 5(1), 103–110.

[3] Mellit, A., Tina, G. M., & Kalogirou, S. A. (2018). Fault detection and diagnosis methods for photovoltaic systems: A review. Renewable and Sustainable Energy Reviews, 91(April), 1–17.

[4] Liu, R., Yang, B., Zio, E., & Chen, X. (2018). Artificial intelligence for fault diagnosis of rotating machinery: A review. Mechanical Systems and Signal Processing, 108, 33–47.

[5] Song, L., Wang, H., & Chen, P. (2018). Vibration-Based Intelligent Fault Diagnosis for Roller Bearings in Low-Speed Rotating Machinery. IEEE Transactions on Instrumentation and Measurement, 67(8), 1887–1899.

[6] Lee, W. K., Cheong, D. Y., Park, D. H., & Choi, B. K. (2020). Performance Improvement of Feature-Based Fault Classification for Rotor System. International Journal of Precision Engineering and Manufacturing,

[7] T. Aydmj and R. P. W. Duin, "Pump failure determination using support vector data description," in Advances in Intelligent Data Analysis (Lecture Notes in Computer Science). Berlin, Germany: Springer, 1999, pp. 415–425.

[8] Sorsa, T., Koivo, H. N., & Koivisto, H. (1991). Neural Networks in Process Fault Diagnosis. IEEE Transactions on Systems, Man and Cybernetics, 21(4), 815–825.

[9] T. Ince, S. Kiranyaz, L. Eren, M. Askar, and M. Gabbouj, "Real-time motor fault detection by 1-D convolutional neural networks," IEEE Trans. Ind. Electron., vol. 63, no. 11, pp. 7067–7075, Nov. 2016.

[10] Eren, L. (2017). Bearing fault detection by one-dimensional convolutional neural networks. Mathematical Problems in Engineering, 2017.

[11] Simonyan, Karen, and Andrew Zisserman. "Very deep convolutional networks for large-scale image recognition." arXiv preprint arXiv:1409.1556 (2014).

[12] Szegedy, C., Liu, W., Jia, Y., Sermanet, P., Reed, S., Anguelov, D., ... & Rabinovich, A. (2014). Going deeper with convolutions. arXiv 2014. arXiv preprint arXiv:1409.4842, 1409.

[13] He, K., Zhang, X., Ren, S., & Sun, J. (2016). Deep residual learning for image recognition. In Proceedings of the IEEE conference on computer vision and pattern recognition (pp. 770-778).

[14] Salamon, J., & Bello, J. P. (2017). Deep Convolutional Neural Networks and Data Augmentation for Environmental Sound Classification. IEEE Signal Processing Letters, 24(3), 279–283.

[15] Wen, L., Li, X., Gao, L., & Zhang, Y. (2018). A New Convolutional Neural Network-Based Data-Driven Fault Diagnosis Method. IEEE Transactions on Industrial Electronics, 65(7), 5990–5998.

[16] Davis, S. Mermelstein, P. (1980) Comparison of Parametric Representations for Monosyllabic Word Recognition in Continuously Spoken Sentences. In IEEE Transactions on Acoustics, Speech, and Signal Processing, Vol. 28 No. 4, pp. 357-366

[17] X. Huang, A. Acero, and H. Hon. Spoken Language Processing: A guide to theory, algorithm, and system development. Prentice Hall, 2001.

[18] Xu, M., & Marangoni, R. D. (1994). Vibration analysis of a motor-flexible coupling-rotor system subject to misalignment and unbalance, part I: theoretical model and analysis. Journal of sound and vibration, 176(5), 663-679.

[19] Muszynska, A., & Goldman, P. (1995). Chaotic responses of unbalanced rotor/bearing/stator systems with looseness or rubs. Chaos, Solitons & Fractals, 5(9), 1683-1704.

[20] Wang, S., Huang, W., & Zhu, Z. K. (2011). Transient modeling and parameter identification based on wavelet and correlation filtering for rotating machine fault diagnosis. Mechanical systems and signal processing, 25(4), 1299-1320.

[21] McFee, Brian, Colin Raffel, Dawen Liang, Daniel PW Ellis, Matt McVicar, Eric Battenberg, and Oriol Nieto. "librosa: Audio and music signal analysis in python." In Proceedings of the 14th python in science conference, pp. 18-25. 2015.

[22] Buitinck, L., Louppe, G., Blondel, M., Pedregosa, F., Mueller, A., Grisel, O., Niculae, V., Prettenhofer, P., Gramfort, A.,

Grobler, J., Layton, R., Vanderplas, J., Joly, A., Holt, B., & Varoquaux, G. (2013). API design for machine learning software: experiences from the scikit-learn project. 1–15.

[23] Tae-Young Jang, Kyu-Sung Kim and Young-Hyo Kim. (2018). Altered Gravity and Immune Response. The KJAsEM, Vol 28, No 1, pp 6–8.

[24] McClellan, James H., Ronald W. Schafer, and Mark A. Yoder. Signal processing first. 2003.

[25] Müller, Meinard. Fundamentals of music processing: Audio, analysis, algorithms, applications. Springer, 2015.

[26] Alex Krizhevsky, Sutskever, Ilya and Hinton Geoffrey E. (2014). ImageNet Classification with Deep Convolutional Neural Networks, Advances in Neural Information Processing Systems 25, pp 1097-1105

[27] Iandola, F. N., Han, S., Moskewicz, M. W., Ashraf, K., Dally, W. J., & Keutzer, K. (2017). 50 X Fewer Parameters and < 0 . 5Mb Model Size. Iclr, pp 1–13.